\documentclass[12pt,preprint]{aastex}

\shorttitle{Candidate water-maser-emitting planetary nebulae}
\shortauthors{G\'omez et al.}

\begin{document}

\title{Radio interferometric observations of candidate water-maser-emitting
  planetary nebulae}

\author{Jos\'e  F. G\'omez\altaffilmark{1}, Olga
  Su\'arez\altaffilmark{2}, 
  Yolanda G\'omez\altaffilmark{3},  Luis F. Miranda\altaffilmark{1}, 
Jos\'e M. Torrelles\altaffilmark{4},
  Guillem Anglada\altaffilmark{1}, \'Oscar Morata\altaffilmark{5}} 

\altaffiltext{1}{Instituto de Astrof\'{\i}sica de Andaluc\'{\i}a,
  CSIC, Apartado 3004, E-18080 Granada, Spain; e-mail: jfg@iaa.es; lfm@iaa.es; guillem@iaa.es}
\altaffiltext{2}{Laboratoire Hippolyte Fizeau, Universit\'e de Nice,
  Parc Valrose, 06108 Nice cedex 02, France; e-mail: olga.suarez@unice.fr}
\altaffiltext{3}{Centro de Radioastronom\'{\i}a y Astrof\'{\i}sica,
  UNAM, Campus Morelia, Apdo. Postal 3-72, Morelia, Michoac\'an 58089,
Mexico; e-mail: y.gomez@astrosmo.unam.mx}
\altaffiltext{4}{Instituto de Ciencias de Espacio (CSIC)-IEEC,
  Facultat de F\'{\i}sica, Planta 
  7a, Universitat de Barcelona, Mart\'{\i} i Franques, 08028
  Barcelona, Spain; e-mail: torrelles@ieec.frc.es}
\altaffiltext{5}{Laboratorio de Astrof\'{\i}sica Espacial y F\'{\i}sica
  Fundamental, INTA, Apartado 50727, E-28080
    Madrid, Spain}

\begin{abstract}
We present Very Large Array (VLA) observations of H$_2$O and OH masers,
as well as
radio continuum emission at 1.3 and 18 cm toward three sources
previously cataloged 
as planetary nebulae (PNe) and in which single-dish detections of H$_2$O
masers have been reported: IRAS 17443$-$2949, IRAS 17580$-$3111, and IRAS
18061$-$2505. Our goal was to unambiguously confirm their nature as
water-maser-emitting PNe, a class of objects of which only two
bona-fide members were previously known. We detected and mapped
H$_2$O maser emission toward all three
sources, while OH maser emission is detected in IRAS 17443$-$2949 and IRAS
17580$-$3111 as well as in other two objects within the observed
fields: IRAS 17442$-$2942 (unknown nature) and IRAS 17579$-$3121 (also
cataloged as a possible PN). We found radio
continuum emission associated only with IRAS 
18061$-$2505. Our results confirm IRAS 18061$-$2505 as the third known
case of a PN associated with H$_2$O maser emission. The three known
water-maser-emitting PNe have clear bipolar morphologies, which
suggests that water maser emission in these objects is related to
non-spherical mass-loss episodes. 
We speculate that these bipolar water-maser-emitting PNe
would have ``water-fountain'' Asymptotic Giant Branch (AGB) and post-AGB stars as their precursors.
A note of caution is given for other objects that have been classified
as OHPNe
(objects with both OH
maser and radio continuum emission, that could be extremely young PNe)
based on single-dish observations,
since interferometric data of both OH masers and continuum are
necessary for a proper identification as members of this class.
\end{abstract}


\section{Introduction}

Maser emission from different molecules is found in the energetic
environments of several types of astronomical objects, such as evolved
stars, young stellar objects, or active galactic nuclei \citep{Eli92}.
Masers have proven to be a very powerful tool to study the morphology
and kinematics of the gas in these environments, since their high
surface brightness makes them excellent targets for radio
interferometric observations, including the use of Very Long Baseline
Interferometry with angular resolutions under 1 milliarcsecond
\cite[e.g.,][]{Tor02, Bob05}.  Therefore, they allow  to study the
inner few AU in young stellar objects and evolved stars, or 1 pc
around active galactic nuclei.

Maser emission from SiO, OH, and H$_2$O are common in the
circumstellar envelopes of oxygen-rich 
evolved stars. In the particular case of water masers, it was
thought for some time that they disappear soon after ($\leq 100$
years) the Asymptotic Giant Branch (AGB) mass-loss stops, so they were not expected to be
associated with PNe \citep{Lew89, Gom90}. However, \citet{Mir01}
reported the source K3-35 as the first confirmed case of a PN
associated with water maser emission (hereafter 
H$_2$O-PN). Later, during a search for this type of masers
toward 26 additional PNe, a new detection was found
toward IRAS 17347$-$3139 \citep{DG04}. In these two cases, the
association of the 
maser emission with the radio continuum emission from the ionized PN
was confirmed via high-resolution interferometric observations using
the Very Large Array (VLA).

Recently, \citet*{Sua07} carried out a single-dish survey for water
masers in post-AGB stars and PNe selected by their IRAS colors in the
atlas of \citet{Sua06}. In that survey, water maser emission was
found toward three objects previously cataloged as PNe
\citep{Rat90,Sua06}: IRAS 17443$-$2949, IRAS 17580$-$3111, and IRAS
18061$-$2505. However, given the relatively large angular separation between
the reported peak position of the 
radio continuum emission \citep{Rat90} and the infrared sources in the
region, \citet{Sua07} suggested that IRAS 17580$-$3111 is probably
not a PN.

In this paper we present VLA observations of maser
emission of H$_2$O and OH molecules, as well as radio continuum
emission, toward IRAS 17443$-$2949, IRAS 17580$-$3111, and IRAS
18061$-$2505.  Our main
goal was to confirm the association of the water maser emission with
these candidates to PNe. It is also interesting to check whether these
masers are 
pumped within a few tenths of AU from the central star, as seen in
IRAS 17347$-$3139 \citep{DG04} and the inner region of K3-35
\citep{Mir01}, or they can also be pumped in shocked outer regions of the PN
lobes, as in K3-35 \citep{Mir01}.  With the evidence provided in this
paper, we can ascertain that IRAS 18061$-$2505 is the third confirmed
case of an H$_2$O-PN.

\section{Observations}

We observed two OH maser transitions at $\simeq 1.6-1.7$ GHz and  the
$6_{16}-5_{23}$ transition of the water molecule at $\simeq 22$ GHz, as well as
radio continuum emission at both $\simeq 1.7$ and 22 GHz, 
toward IRAS 17443$-$2949, IRAS 17580$-$3111, and IRAS 18061$-$2505, using the
VLA of the National Radio 
Astronomy Observatory (NRAO)\footnote{ The National Radio Astronomy
  Observatory is a facility of the National Science Foundation
  operated under cooperative agreement by Associated Universities,
  Inc.}.
The details of the observations are listed in Tables
\ref{tab.obslogline}, \ref{tab.obslogcont}, and \ref{tab.sourcedata}. 
In all cases, both right and left circular polarization were observed. 
We note that a continuum dataset at 1.7 GHz, 
with a bandwidth of 25 MHz,  was obtained
simultaneously with the OH line at 1665 MHz. However, these
continuum data 
were not usable due to radio frequency
interference. An alternative continuum dataset at that frequency was
obtained by 
averaging line-free channels in the OH spectral data at 1665 MHz.
The H$_2$O maser and radio
continuum observations at
$\simeq 22$ GHz were also carried out simultaneously.

The data
were calibrated and processed using the Astronomical 
Image Processing System (AIPS) of the NRAO. 
Maps were obtained with robust weighting (robust parameter = 0) of
visibilities \citep{Bri95}, using the AIPS 
task IMAGR. When a maser
line was detected, self-calibration was applied on the channel with
maximum emission. The self-calibration solutions were then applied to
all channels and, in the case of the H$_2$O masers, these solutions
were also applied to the simultaneous radio continuum observations at
22 GHz. For the OH data at 1665 MHz in IRAS 17443$-$2949,
self-calibration was applied to the continuum emission from GAL 359.23$-$00.82 
\citep[The Mouse Nebula, detected by][]{Yus87}, which fell within the primary
beam; this continuum 
emission was extracted by integrating 116 channels from the dataset.
When present, continuum emission was subtracted from  line data using the AIPS
task UVLIN. 
Some line data sets had to be Hanning-smoothed in frequency (see Table
\ref{tab:result_line}), to
alleviate spectral ripples due to the Gibbs phenomenon.

\section{Results}

\subsection{Detected emission}

Our results are summarized in Tables \ref{tab:result_line} and
\ref{tab:result_cont}. 
We have detected and mapped H$_2$O maser emission toward all three
target sources. OH maser emission at 1612 MHz was detected in IRAS
17443$-$2949 and IRAS 17580$-$3111, but not in IRAS 18061$-$2505. OH
emission at 1665 MHz was only detected in IRAS 17443$-$2949. As for
the radio continuum emission, it was only detected associated with IRAS
18061$-$2505, at both 
1.7 and 22 GHz .

In addition to this, OH maser emission at 1612 MHz was detected toward two
additional sources that lie within the field of view of the
observations (primary beam $\simeq 28'$): IRAS 17442$-$2942 (within
the primary beam of the 
observations toward IRAS 17443$-$2949), and IRAS 17579$-$3121 (within
the primary beam from IRAS 17580$-$3111). These additional two sources
were well outside the corresponding primary beam of the 22 GHz
observations ($\simeq 2'$ size), so it is not possible to know whether they are
associated with H$_2$O maser emission.

In the following sections we
present the spectra and maps of the maser emission toward all five sources. 

\subsection{Individual objects: target sources}

\subsubsection{IRAS 17443$-$2949}

This source has been classified as PN, based on the detection of radio
continuum emission \citep{Rat90}, with a flux density $S_\nu {\rm (6
  cm)}\simeq 0.9$ mJy. Water maser spectra obtained by \citet{Sua07} showed two
well-defined spectral components separated by 2 km s$^{-1}$, centered at
$\simeq -5$ km s$^{-1}$. 
Maser emission of OH at 1612 MHz was previously
reported by \citet{Zij89}, with a single peak at $\sim -15$ km s$^{-1}$. These authors
also point out that OH emission coincides, within the errors, with the
radio continuum emission.

\begin{figure}
\includegraphics[scale=0.8]{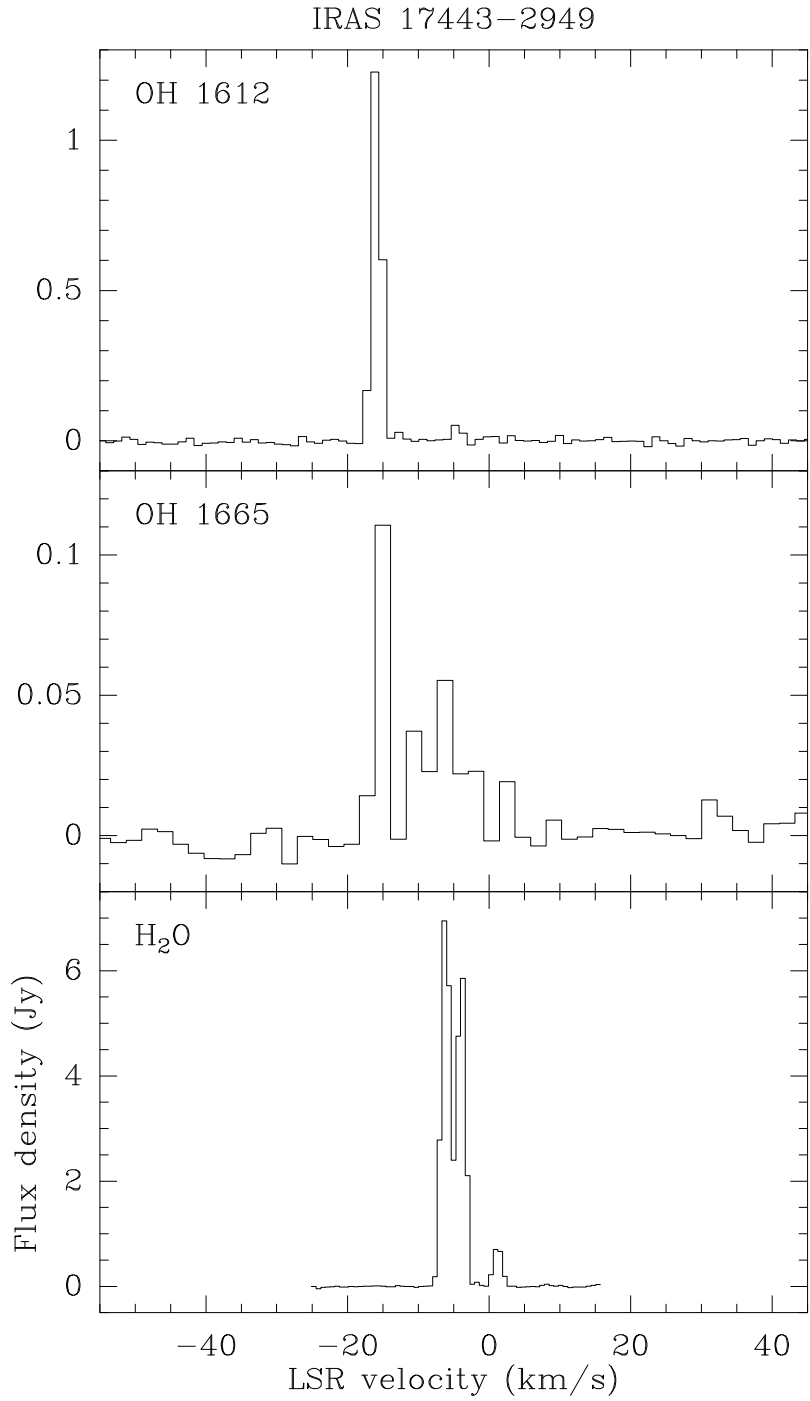}
\caption{OH and H$_2$O maser spectra toward IRAS 17443$-$2949}.
\label{fig:spec17443} 
\end{figure}

Our H$_2$O maser spectrum shows four spectral components above the
noise level (Table \ref{tab:result_line} and
Fig. \ref{fig:spec17443}), although the component at $-1.7$ km
s$^{-1}$ is significantly weaker than the other three and it is difficult to
distinguish in Fig. \ref{fig:spec17443}. The two most intense
components correspond to the ones detected in the single-dish observations of
\citet{Sua07}. The H$_2$O maser components are closely clustered in space
(Table \ref{tab:result_line}), 
within a region of $\simeq 80$ 
milliarcsec. No obvious spatio-kinematical distribution is seen, and
therefore it is not possible to determine what kind of structure the
water masers are tracing.

\begin{figure}
\includegraphics[scale=.8,angle=-90]{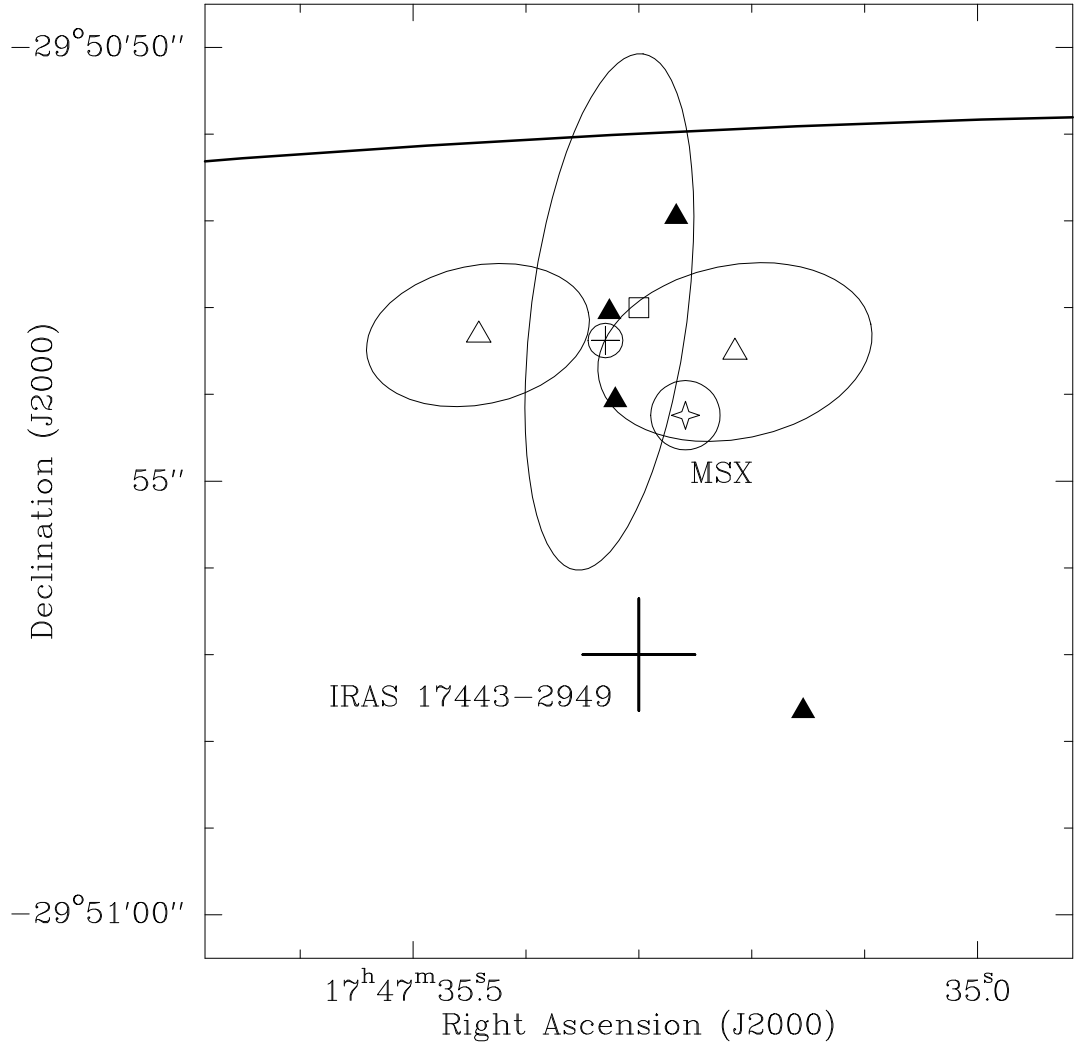}
\caption{Positions of maser components and infrared sources in the IRAS
  17443$-$2949 region. Open and filled triangles represent the OH maser
  components at 
   1612 MHz and 1665 MHz, respectively. The small cross
  represents the mean position of the H$_2$O maser components. The large cross marks
  the position of IRAS
  17443$-$2949 in the IRAS point source catalog. The four-pointed star
  is the position of the MSX infrared source MSX6C G359.4428-00.8398. 
Ellipses represent the estimated absolute positional error of each
source and maser component 
(for OH-1665-MHz only the error ellipse of the
component with best positional accuracy is plotted). The thick curve is part of the error ellipse of
the IRAS source position (semiaxes of the error ellipse: $34''\times
6''$, p.a. $= 93^\circ$). The open square is the position of the
1612-MHz OH maser reported by \citet{Zij89}, for which we did not plot
an error ellipse.
}
\label{fig:map17443all} 
\end{figure}

We have detected both OH lines at 1612 and 1665 MHz in this object (for the first
time in the case of the latter transition). The spectrum of the 1612-MHz line
shows a double peak, although very asymmetric, with the component at
$V_{\rm LSR}
\simeq -16.1$ km s$^{-1}$ being $\sim 20$ times stronger than the one
at $\simeq -4.8$ km s$^{-1}$  (Fig. \ref{fig:spec17443}). The OH spectrum at 1665 MHz shows also a peak
close in velocity to that at 1612 MHz, with at least three 
additional spectral features
covering the velocity range in which H$_2$O masers are also detected.
 The two components of the OH 1612-MHz spectrum
appear significantly separated in space ($\simeq 3''$). The H$_2$O
masers lie midway between these two OH 1612-MHz 
components (Fig. \ref{fig:map17443all}). On the other hand, given the large relative positional errors for the OH 1665 MHz
emission, it is not possible to determine its real spatial
distribution.

Fig. \ref{fig:map17443all} shows the possible infrared counterparts
for the object powering the masers. We have checked both the MSX
\citep{msx} and
2MASS \citep{2mass} catalogs, in the mid- and near-infrared ranges,
respectively, to search for a counterpart of the IRAS source that
could be the exciting source of the maser emission. The infrared
point source that is closer to the maser positions is MSX6C G359.4428-00.8398,
which also lies 
within the error ellipse of IRAS 17443$-$2949. Its position is consistent,
within the errors, with the OH 1665-MHz emission and one of the OH 1612-MHz
components.

We did not detect radio continuum emission, either at 1.7 or 22
GHz (Table \ref{tab:result_cont}),
that could be associated with the maser emission. 
The radio continuum source at 4.9 GHz reported by \citet{Rat90} is $\simeq 9''$
away from the position of the water masers and $10''$ from the MSX
source.  
We
note that the beam of the \citet{Rat90} observations was $\simeq
5''-10''$, although we would expect their absolute positional accuracy
to be somewhat better than that. We think it is unlikely that the masers
(specially those of H$_2$O, with a better absolute positional
accuracy, $\simeq 0\farcs 2$) are
associated with the reported radio continuum emission. Therefore, we
do not have enough evidence to support that the maser emission we
detected may be pumped by a PN. 

The identification of IRAS 17443$-$2949
itself as a PN is also questionable.
 Recently, \citet{Gar07} suggested
that this source is still in the AGB phase, given its high variability in
the infrared and the presence of strong amorphous silicate absorption
features in its infrared spectrum. Its AGB status, together with the
presence of OH and H$_2$O emission, and its unlikely association with
radio continuum emission, lead us to classify this source as an
OH/IR star, rather than pertaining to the OHPN class defined by
\citet{Zij89}.

\subsubsection{IRAS 17580$-$3111}

This source has also been classified as a PN, with detected radio
continuum emission at a level of $S_\nu {\rm (6  cm)}\simeq 2.5$ mJy
\citep{Rat90}. \citet{Sua07} detected
water maser emission, dominated by a component at $\sim 21$ km
s$^{-1}$. OH maser emission at 1612 MHz was
detected by \citet{Zij89}, showing at least four spectral
components. 
These authors pointed out the need
of VLA observations to determine whether more than one source within
the beam of their observations could be contributing to the OH
spectrum.

\begin{figure}
\includegraphics[scale=0.8]{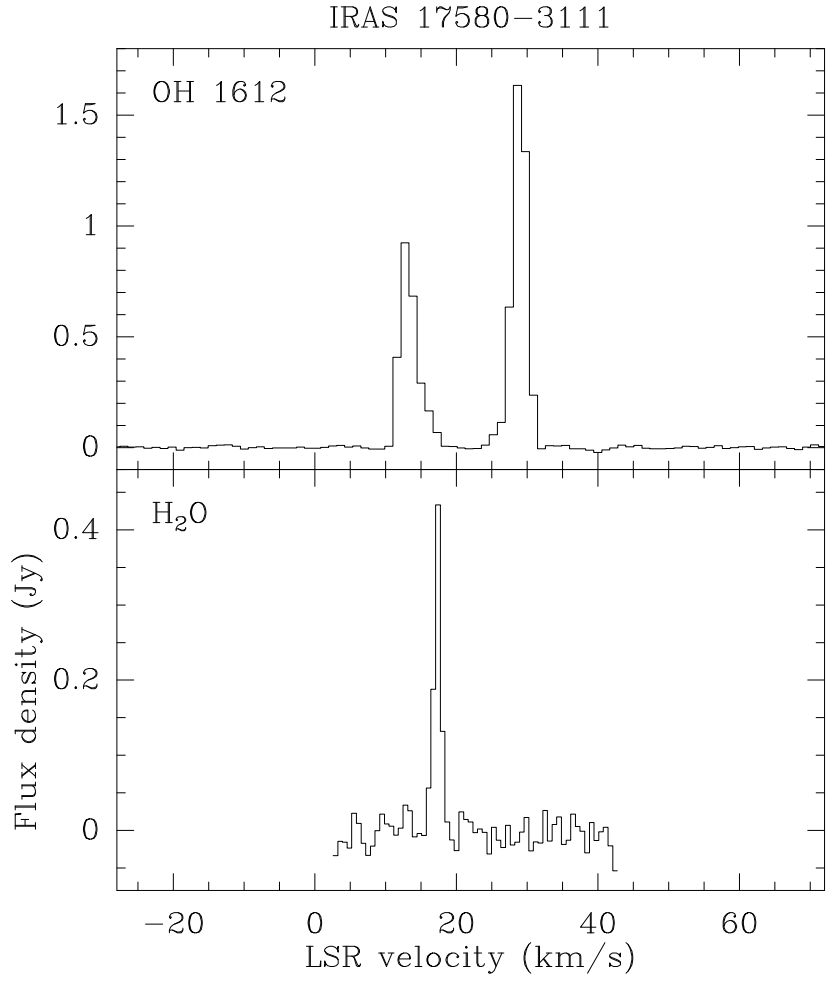}
\caption{OH and H$_2$O maser spectra toward IRAS 17580$-$3111.}
\label{fig:spec17580} 
\end{figure}

Our H$_2$O maser data shows only one component
(Fig. \ref{fig:spec17580}), at 17.4 km 
s$^{-1}$.
The OH 1612-MHz spectrum shows a typical two-peaked spectrum
(Fig. \ref{fig:spec17580}), whose components coincide in 
velocity with two of the four components detected by
\citet{Zij89}. The remaining two components seen by these authors may be
associated with the source IRAS 17579$-$3131 (see
sec. \ref{sec:I17579}), located $\simeq 11'$ away from IRAS
17580$-$3111, although the former is outside the half-power beam of
the telescope ($\simeq 12\farcm 6$)  in
their single-dish observations. The positions of the two OH components
we detected are consistent with each other, within their relative
errors (Table \ref{tab:result_line}). No OH maser at 1665 MHz, nor
radio continuum emission at either 1.7 or 22 GHz was
detected (Tables \ref{tab:result_line}, \ref{tab:result_cont}). 

\begin{figure}
\includegraphics[scale=.8,angle=-90]{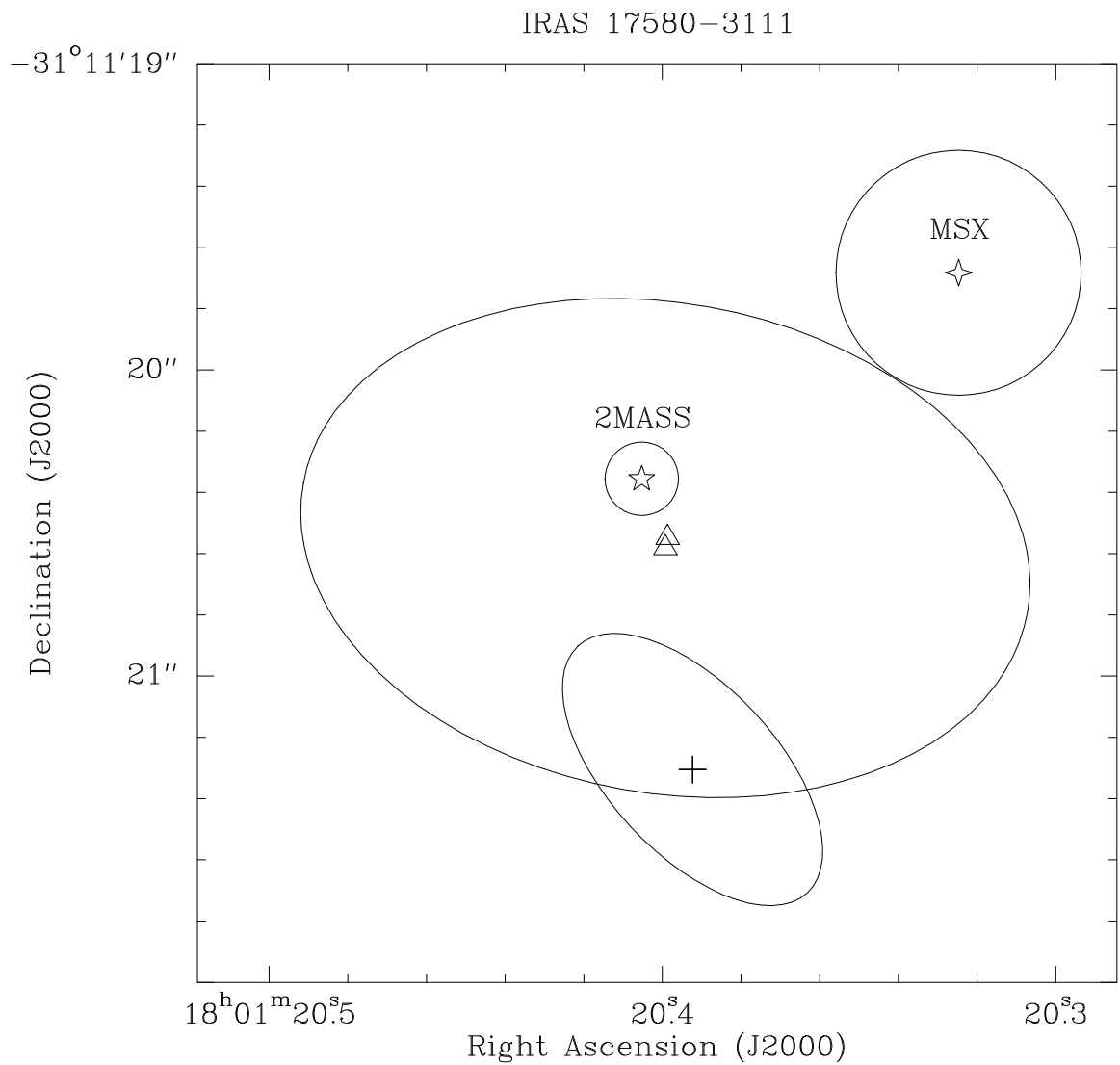}
\caption{Positions of maser components and infrared sources in the IRAS
  17580$-$3111 region. The cross and open triangle represent the
  H$_2$O and OH-1612-MHz maser components, respectively. The four-pointed star
  is the position of the MSX infrared source MSX6C
  G359.7798-04.0728. The five-pointed star marks the position of the
  source 2MASS J18012040-3111203. 
Ellipses represent the estimated absolute positional error of each
component and source. The whole area shown is within the error ellipse
of IRAS
  17580$-$3111 [located at R.A.(2000) = $18^h01^m19.6^s$, Dec.(2000) =
  $-31^\circ 11'20''$. Error ellipse semiaxes: $38''\times 7''$, p.a. = $91^\circ$].
}
\label{fig:map17580all} 
\end{figure}

Fig. \ref{fig:map17580all} shows the position of all maser
components and infrared point sources in their neighborhood. There is
a mid-infrared source (MSX6C
  G359.7798-04.0728) and a near infrared one (2MASS J18012040-3111203)
  close to the masers. Both infrared sources as well as the H$_2$O and
  OH maser components lie within the error ellipse of IRAS
  17580$-$3111. The IRAS, MSX, and 2MASS fluxes are, in principle, 
compatible with belonging
  to the same object. However, the error ellipses quoted in the 
 MSX and 2MASS catalogs
  do not intersect, so they might actually 
 be tracing different sources. Given its position, the object
  2MASS J18012040-3111203 is the most likely candidate to be the
  powering source of the maser emission.

Regarding the nature of the object, as already pointed out by
\citet{Sua07}, the radio continuum source reported by \citet{Rat90} is
more than $30''$ away from the infrared sources and the maser emission of
both OH and H$_2$O, and
therefore, it is not associated with them. The identification of IRAS
  17580$-$3111 as a PN on the basis of the presence of radio continuum
  emission is not well justified. \citet{Gar07} also presented
  infrared spectroscopy toward this IRAS source; its spectrum and its low
  variability in the infrared led these authors to classify IRAS
  17580$-$3111 as a 
  post-AGB star. 

\subsubsection{IRAS 18061$-$2505}

This source was first cataloged as a PN by
\citet{Mac78} (object Mac 1-10). It shows a bipolar morphology with
two well-defined lobes seen in the H$_\alpha$ image \citep{Sua06}, with
a total size of $\simeq 46''$. Images of the continuum close in
wavelength to the
H$_\alpha$ line, as well as two-dimensional spectroscopy (Su\'arez et
al. 2008, in preparation) clearly show
that the lobes are ionized. This source also shows radio continuum
emission at 1.4 GHz in 
the NRAO/VLA Sky Survey \citep{Con98}, with a flux density of $3.5 \pm
1.0$ mJy, at R.A.(J2000) $=18^h 09^m 12\fs 9$, Dec(J2000) $= -25^\circ
04' 27''$ (positional error $\simeq 7''$). \citet{Sua07} detected
water maser emission, with three highly-variable components.

\begin{figure}
\includegraphics[scale=0.8]{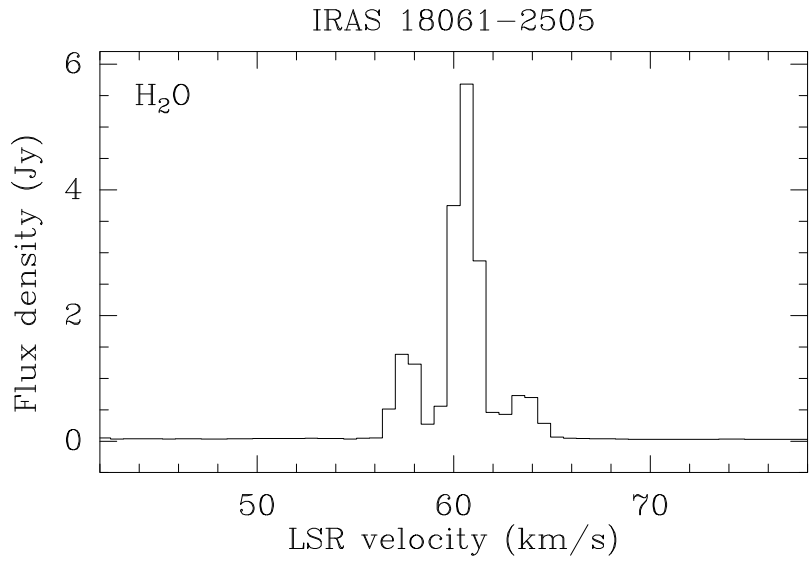}
\caption{H$_2$O maser spectrum toward IRAS 18061$-$2505.}
\label{fig:spec18061} 
\end{figure}

We detected water maser emission toward IRAS 18061$-$2505
(Fig. \ref{fig:spec18061}). No OH maser emission at either 1612 or 1665 MHz
was
detected. The H$_2$O maser emission exhibits three distinct
spectral features, at $\simeq 57$, 61, and 64 km s$^{-1}$. We also
detected unresolved 
radio continuum emission at 1.7 and 22 GHz from the ionized region
associated with this source (Table \ref{tab:result_cont}), whose position is
consistent with that 
given by \citet{Sua06} for the central
region of the cataloged PN.
We did not detect continuum emission in the large ionized lobes seen in the H$_\alpha$
image. The dynamic range of our radio continuum map at 22 GHz is $\simeq 400$
(rms $\simeq 90$ $\mu$Jy beam$^{-1}$), and the emission from the extended lobes must be
below the sensitivity level imposed by this dynamic 
range. The positions of the infrared sources 2MASS J18091242-2504344 and MSX6C
G005.9737-02.6115 are consistent, within the errors, with the position
of the radio continuum source.

The nominal positions
of these maser features are all within 50 mas 
\citep[65 AU assuming a distance of 1.3 kpc,][]{Pre88}
from one another and from the radio continuum
emission, which confirms that IRAS 18061$-$2505 is the third PN known
to harbor water masers. We note that no water maser emission has been
found associated with the extended lobes of the PN. It still remains
to be determined whether the water maser emission found associated with the
central source could be tracing the inner part of a jet or a
circumstellar structure, such as a disk or torus. However, the latter
possibility seems likely, given that water masers in the other two known
H$_2$O-PNe (Miranda et al. 2001; de Gregorio-Monsalvo et al. 2004;
Uscanga et al. 2008, in preparation)
seem to preferentially 
trace toroidal structures. A
confirmation of the spatial distribution of water masers in IRAS 18061$-$2505
  would only be possible  
with a more extended VLA configuration 
or with VLBI networks. 

Using our radio continuum data, we can make a rough estimate of the
spectral index of $\alpha = 0.92\pm 0.13$ 
($S_\nu\propto \nu^\alpha$) between 1.4 and 22 GHz. This index is
similar to the one found in IRAS 17347$-$3139, $\alpha = 0.79\pm
0.04$, between 4.3 and 8.9 GHz \citep{Gom05}, whose rising
spectral energy distribution beyond 10 GHz was suggested to be a
signature of youth. Simultaneous observations of radio continuum at
different frequencies would be necessary to obtain a proper estimate
of the spectral index and the turnover frequency of IRAS 18061$-$2505,
to derive physical properties such as its emission measure and
electron density.

We also note that, among the three confirmed PNe with water masers, this is
the only one with no detectable OH maser emission, and it is also the one
with the largest angular size \citep[cf:][]{Mir01,DG04,Sua06}.

\subsection{Individual objects: other sources in the fields}

\subsubsection{IRAS 17442$-$2942}
\label{sec:I17442}

This source is cited in the SIMBAD database as associated to the
gamma-ray source 3EG J1744-3011 in the third EGRET catalog
\citep{Har99}, or 2EG J1747-3039 in their second catalog
\citep{Tho95}. 
However, the IRAS position is 
1.3 and 0.7 degrees, away from the positions in the third and second
EGRET catalogs, respectively. This means that the IRAS source is
outside the error ellipse of the gamma-ray source (0.32 and 0.2
degrees, depending on the catalog version). Therefore, we think that
the identification of the IRAS source with the gamma-ray source is
incorrect. 

\begin{figure}
\includegraphics[scale=0.8]{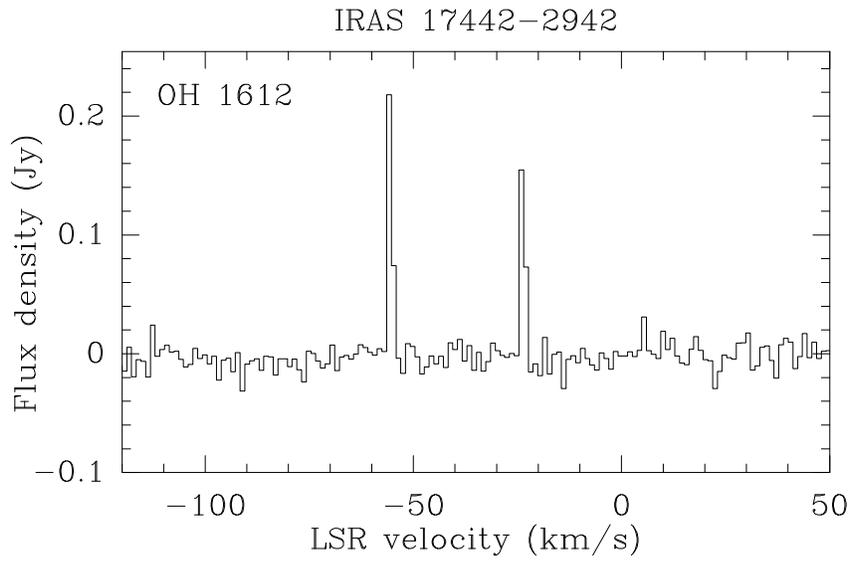}
\caption{OH maser spectrum toward IRAS 17442-2942.}
\label{fig:spec17442} 
\end{figure}

We have detected OH 1612-MHz emission associated with IRAS
17442$-$2942 (Fig. \ref{fig:spec17442} and Table \ref{tab:result_line}), which
was within
the primary beam of our observation toward IRAS 17443$-$2949. There are
two well-defined velocity components, separated by $\simeq 30$ km
s$^{-1}$. No OH maser at 1665 MHz nor radio continuum emission at
18 cm has been detected. This source is well outside the primary beam of the
H$_2$O observations, so it is not possible to determine whether there
is any H$_2$O maser emission associated with this object.

\begin{figure}
\includegraphics[scale=.8,angle=-90]{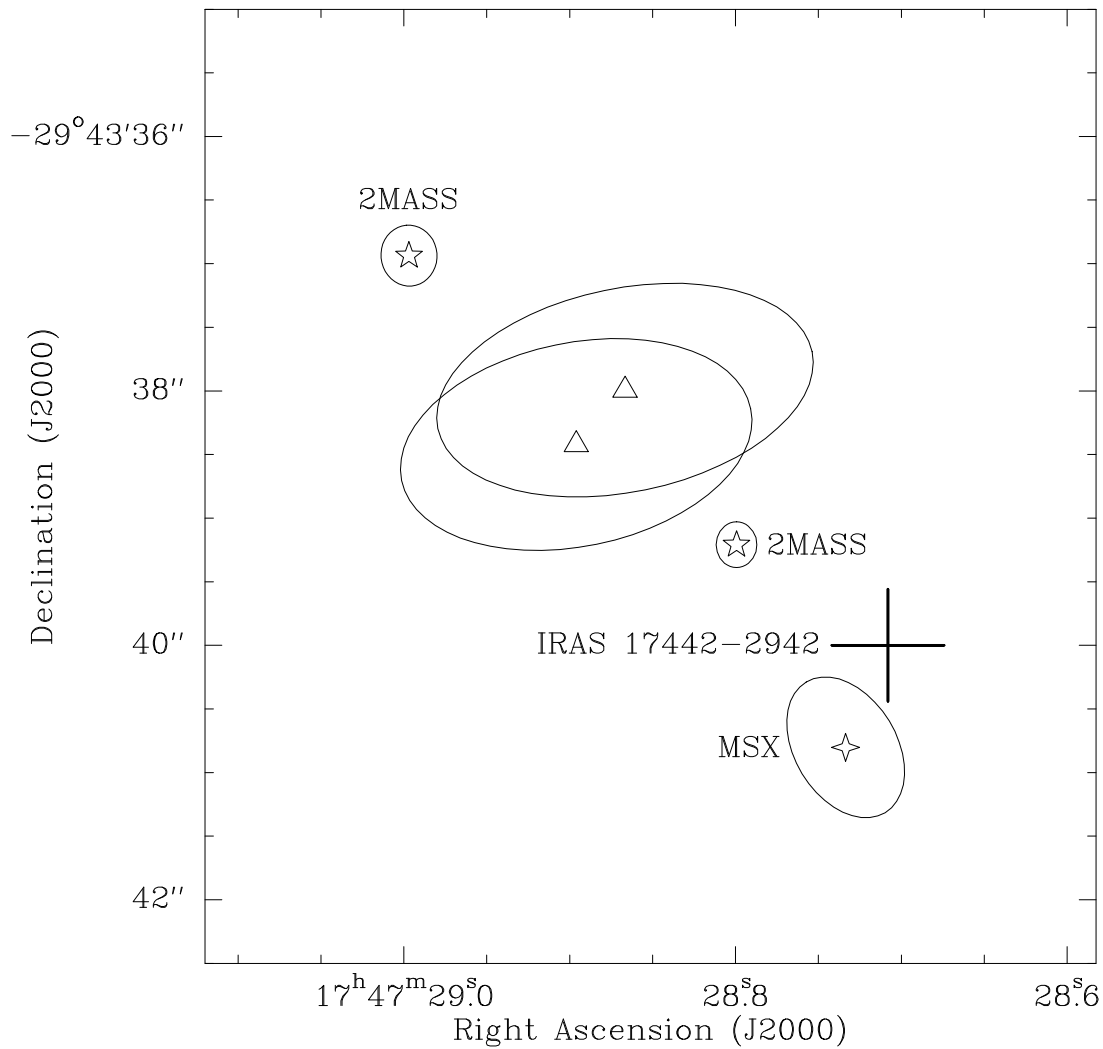}
\caption{Positions of maser components in and infrared sources in the IRAS
  17442$-$2942 region. The open triangles mark the location of the OH
  maser components at 1612 MHz. The cross marks
  the position of IRAS
  17442$-$2942 in the IRAS point source catalog (semiaxes of error
  ellipse: $50''\times 7''$, p.a. $= 92^\circ$). The four-pointed star
  is the position of the MSX infrared source MSX6C
  G359.5336-00.7573. The five-pointed stars mark the position, in
  order of increasing right ascension, of sources 2MASS
  J17472879-2943392 and 2MASS J17472898-2943369.
Solid ellipses represent the estimated absolute positional error of each
component and source. 
}
\label{fig:map17442all} 
\end{figure}

Both OH components are located midway between two infrared sources of
the 2MASS catalog, 2MASS J17472879-2943392 and 2MASS 17472898-2943369
(Fig. \ref{fig:map17442all}), 
but it is not possible to determine whether either of
them is pumping the maser. With the scarce information available,
 we do not know the evolutionary
stage of the pumping source of the maser emission in this region.

\subsubsection{IRAS 17579$-$3121}

\label{sec:I17579}
This source is also listed as a PN by \citet{Rat90}, although 
it shows an optical spectrum typical of a post-AGB star
\citep{Sua06}. The radio continuum source reported by \citet{Rat90}
has a flux density $S_\nu({\rm 6cm})\simeq 0.8$
mJy and is $\simeq 12''$ from
the possible infrared counterparts of this evolved object (2MASS
J18011337-3121566 and MSX6C G359.6129-04.1379), although all these
sources are within the error ellipse of IRAS 17579$-$3121. \citet{Sly97} reported a
single-dish 
detection of OH maser emission at 1667 MHz, although it may actually arise
from IRAS 17580$-$3111, since the position of this source 
is close to the half-power level of the
telescope beam
in their observations.

\begin{figure}
\includegraphics[scale=0.8]{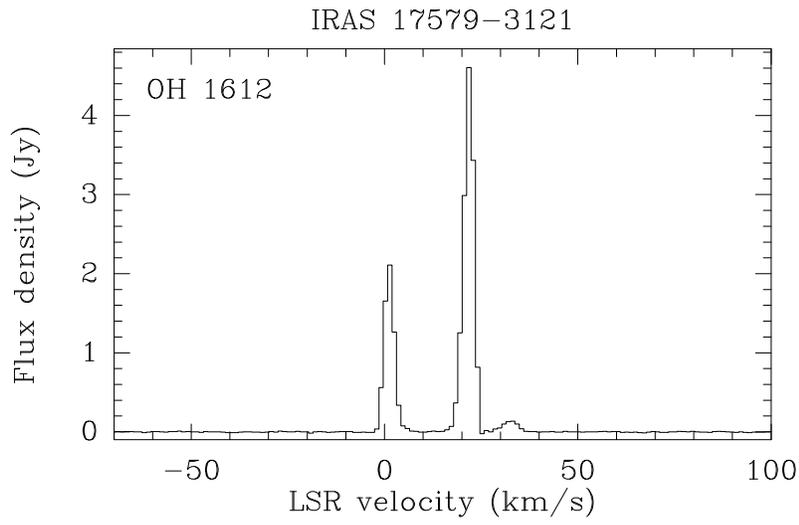}
\caption{OH maser spectrum toward IRAS 17579$-$3121.}
\label{fig:spec17579} 
\end{figure}

This source was within the primary beam of our OH observations toward
IRAS 17580$-$3111. We detected OH maser emission at 1612 MHz, but not at
1665 MHz. It shows three distinct spectral components
(Fig. \ref{fig:spec17579} and Table \ref{tab:result_line}). The two 
strongest components may account for two of the four ones detected in the
single-dish observations of IRAS 17580$-$3111 reported by
\citet{Zij89}. 
All three OH components are coincident, within the errors, with 2MASS
J18011337-3121566. No radio continuum emission at
18 cm has been detected. The source is also well outside the primary
beam of the H$_2$O observations and therefore, it is not possible to
determine its 
association with H$_2$O maser emission. 

\section{Discussion}

\subsection{H$_2$O and OH masers and the evolution of bipolar PNe}

One of the main results in this paper is the confirmation of IRAS 18061$-$2505
as a water-maser-emitting PN. This is the third object that could be
considered a bona fide H$_2$O-PN,
only after K3-35 \citep{Mir01} and IRAS 17347$-$3139
\citep{DG04}. In classical studies of maser emission in evolved
stars, water masers were not expected in stages as late as PNe. However,
since
the number of known cases is growing, it seems interesting to review
the characteristics of H$_2$O-PNe, and how they fit in the general
scheme of late stellar evolution.

Many previous studies of maser emission (of SiO, H$_2$O, and OH
molecules) in evolved stars focused on
the AGB phase, mainly Mira-type and OH/IR stars. OH maser
emission in these objects (and in a significant fraction of post-AGB
stars) typically
shows double-peaked profiles, specially for the 1612 MHz transition, 
with narrow
components separated by $\la 30$ km s$^{-1}$ \citep[e.g.,
][]{Ede88,Sev97}.
These profiles have been 
interpreted 
as arising from the approaching 
and receding sides of an spherically-symmetric, expanding envelope
\citep{Rei77}. This seems to be the case of IRAS
17580$-$3111 or IRAS 17442$-$2942 (Figs. \ref{fig:spec17580} and \ref{fig:spec17442}) in this paper.
In
some cases only one peak is detected, which could be due 
to sensitivity limitations if the spectrum is asymmetric, with
one of the peaks significantly weaker than the other. 
Emission of H$_2$O and SiO tend also to
be double- or single-peaked, with their spectral components
encompassed within the range marked by the two peaks of the OH
emission at 1612 MHz. The presence of different maser transitions was
used by \citet{Lew89} to construct a chronological sequence, in which
masers appear and 
disappear sequentially as the star evolves from the AGB to the PN
phase. However, as noted  by \citet{Lew89}, this chronological
sequence is only valid for the case of 
spherically symmetrical mass-loss, and it will not stand if bipolar
mass loss is present.

The presence of maser spectra not following the regular double (or
single) peak pattern, or with H$_2$O maser emission extending outside the
velocity range 
covered by OH emission, 
are interpreted as evidence of non-spherical mass-loss
processes \citep[e.g.,][]{Gom94,Dea04}. That would be the case of sources IRAS
17443$-$2949, IRAS 18061$-$2505, or IRAS 17579$-$3121 (Figs.
\ref{fig:spec17443}, \ref{fig:spec18061}, and \ref{fig:spec17579}) in
this paper.  
A significant fraction of OH and H$_2$O maser spectra in 
post-AGB stars do show multiple velocity
components \citep{Sev02,Eng02}. 
Specially significant is the case of ``water fountain''
sources (\citealt{Lik88}; see \citealt{Ima07} for a review). These are
late-AGB and post-AGB stars which show H$_2$O maser 
components spanning $\ga 100$ km s$^{-1}$, i.e., significantly
larger than the few tens of km s$^{-1}$ that characterize the
spherical wind in the AGB. Water emission in water fountain
 sources traces highly collimated and symmetrical jets, with
dynamical ages of 
10-100 yr \citep{Sah99,Ima02,Cla04,Ima07et,Ima07,Mor03,Bob07}. These sources
represent the earliest known manifestation of 
non-spherical mass-loss in evolved stars. We also note that, in some
of these sources, in addition to the high-velocity
maser components associated with the jet, there are low-velocity
components located close to the dynamical center of the jet
\citep{Ima07}. These 
components could trace
low-velocity equatorial flows, or expanding toroidal structures.

Focusing again on H$_2$O-PNe, 
a key common property is that  all three
confirmed cases show a clear
bipolar morphology at optical, infrared, and/or radio continuum
\citep{Mir01,DG04,Sua06}. This
could indicate  
that water masers in 
PNe are related to  non-spherical mass-loss phenomena, rather than being
the remnant of the masers excited in the (spherical) AGB wind \citep{Sua07}.
As discussed elsewhere
\citep[e.g.][]{Eng02,Dea07,Sua07}, the presence of water fountains and
H$_2$O-PNe shows that 
H$_2$O masers 
in evolved stars 
could be pumped in two different phases: by spherical winds during the
AGB phase, and
later, by highly collimated winds that turn on close to the end
of the AGB phase for a  particular type of stars (probably the most massive
precursors of PNe). The sources in which H$_2$O masers are excited by these
collimated jets would first appear as water fountains and, following
\citet{Sua07}, we speculate that some of them might evolve to become
H$_2$O-PNe. 

We note here the different distributions and kinematics of H$_2$O maser
emission in water fountains and H$_2$O-PNe. While H$_2$O maser
emission in water fountains 
always traces high-velocity bipolar jets, with additional
low-velocity equatorial structures in some cases \citep{Ima07}, the emission in
H$_2$O-PNe tends to trace compact, low-velocity 
equatorial structures. Only in one
epoch, in the H$_2$O-PN K3-35, water maser emission was detected associated
with a bipolar jet \citep{Mir01}, although with a much lower
line-of-sight velocity ($\le 5$ km s$^{-1}$) than jets
traced by masers in water fountains. Water masers tracing jets have
not been seen in 
subsequent observations of K3-35, neither in IRAS 17347$-$3139
\citep{DG04}, which
suggests that this could be a transient phenomenon in H$_2$O-PNe.

To explain the short dynamical ages of water fountains ($\la 100$ yr), 
\citet{Ima07et} suggested that the high-velocity, collimated jet
drills through the circumstellar envelope 
previously expelled during the AGB phase. Shocks between the jet and the
envelope excite H$_2$O
masers along the jet direction. A possible coeval equatorial flow
could excite water masers along a perpendicular direction. When the
jet reaches the lower
density regions in the outer envelope, water
masers can no longer be pumped along the jet. A jet of $\simeq 100$
km s$^{-1}$ would advance $\simeq 2000$ AU in $\simeq 100$ yr, which is the
typical size for circumstellar envelopes in OH/IR stars. At that point, these
sources would stop showing the characteristic maser emission of water
fountains, although jets could still be active, even during the PN
phase \citep{Vel07}.

If we extrapolate this scenario beyond the point at which 
the high-velocity masers of the water-fountain jets 
turn off, the equatorial
regions are the only ones close enough to the central star and with
the necessary 
high density to eventually support water maser emission. By the time
the central
source starts the ionization of its envelope, masers could be excited
preferentially in equatorial 
regions (circumstellar disks or equatorial flows), as seen in
H$_2$O-PNe. Dense molecular gas has been found
in K3-35 \citep{Taf07}, 
providing the conditions in which molecules like H$_2$O 
can survive in the
ionizing environment of a PN \citep{Taf07}.

Summarizing, we propose the following scenario for water
maser emission during the evolution toward bipolar PNe:
\begin{enumerate}
\item Water-fountain phase. Water masers preferentially trace a
  high-velocity jet and, in some cases, equatorial regions. This phase
  ends when the jet reaches regions with a density too low for water
  masers to be excited.
\item H$_2$O-PN phase. Young bipolar PNe, with its ionized structure
  extending along  the lower-density bipolar cavity previously opened by
  jets. Water masers 
  could be excited mostly 
  in the denser equatorial regions.
\end{enumerate}

Whether there is a significant gap between these two phases, or they
overlap, would probably depend on the mass of the progenitor star.
However, since the progenitors of bipolar PNe are thought to be relatively
massive, their evolution would be fast,
and ionization could start soon after the
onset of bipolar mass-loss. In this case, the two phases mentioned
above could
overlap, and 
 at some point of their evolution they could be PNe with
water-fountain characteristics.  No water-fountain PN has been
confirmed 
so far, but if any such object exists, our evolutionary scenario would
indicate that it should have a high optical extinction (thus making difficult
its identification as a PN) and the mass of its progenitor
star should be in the upper ranges for PNe ($\simeq 8$ M$_\odot$).

We note that, in our scenario, not all water fountains may necessarily
end up as H$_2$O-PNe, if the circumstellar envelope has been
significantly cleared, even in equatorial regions, so that the
conditions to pump water masers are not met anywhere in the
circumstellar envelope
by the time ionization starts. However, we think
it likely that all H$_2$O-PNe have gone through a phase with
collimated jets, probably shown as water fountains.

\subsection{A note on planetary nebulae with OH emission (OHPNe)}

The apparent non-association of water masers with radio continuum
emission in IRAS 17443$-$2949 and IRAS 17580$-$3111 is also worth noting. 
Both sources have been previously classified as PNe on the basis
of presence of radio continuum emission \citep{Rat90}. The detection of OH
emission made \citet{Zij89} to include them in a special class of objects,
OHPNe. This type of objects, showing both OH and radio continuum emission, and
strongly obscured or invisible in the optical, 
were suggested to form an
evolutionary phase immediately before the formation of a full-blown
PN. Therefore, they are potentially key objects to study the early
evolution of PNe, including the processes that may give rise to the different
morphologies observed in these nebulae. 

Our results, using an angular resolution higher than that available in the
data presented by \citet{Zij89}, suggest that in IRAS 17443$-$2949 and IRAS
17580$-$3111, OH and H$_2$O maser emission is not associated with the reported
radio continuum emission and, therefore, they are not proper members of the
OHPN class. We also note  that \citet{Rat90} reported radio continuum
emission toward IRAS 17579$-$3121, and we have detected OH maser emission, although
its angular separation to the radio continuum is too large to be associated,
and thus, 
this source should not be considered an OHPN either.

Therefore, it seems very important to carry high angular and spectral
resolution studies 
of maser and radio continuum emission in all sources that have been proposed as
candidates to being OHPNe, before drawing any further conclusions about their
properties as a group and their relevance as a link between the post-AGB and
PN phase. For some objects, such as K3-35 and IRAS 17347$-$3139,
observations with high
enough angular resolution allow us to include them as proper members of the
OHPN class. Other prospective OHPNe
would probably need more detailed research before being unambiguously
included within this class of objects.

\section{Conclusions}

We have presented VLA observations of OH (1612 and 1665 MHz) and water masers
(22235 MHz) as well as
radio continuum emission at 1.3 and 18 cm toward three possible
water-maser-emitting 
PN: IRAS 17443$-$2949, IRAS 17560$-$3111, and IRAS 18061$-$2505. Our
main conclusions are as follow:

\begin{itemize}
\item We have detected water maser emission toward all target
  sources. OH maser emission at 1612 MHz is found associated with IRAS
  17443$-$2949 and IRAS 17560$-$3111, as well as
  toward two other objects within the observed fields: IRAS
  17442$-$2942 (unknown
  nature) and IRAS 17579$-$3121 (previously cataloged as a possible
  PN). OH maser 
  emission at 1665 MHz is present in IRAS 17443$-$2949. Radio continuum
  emission at  1.3 and 18 cm was detected only 
  toward IRAS 18061$-$2505.
\item The water maser and radio continuum emission in IRAS
  18061$-$2505 is found within a region of 50 mas in size, coincident with the
  central region of the large ($\simeq 46''$ size) PN seen in the optical
  images. This confirms this object as the third known
  water-maser-emitting PN. The three confirmed H$_2$O-PNe have clear
  bipolar morphologies, which suggest that the water maser emission in
  these objects is not the remnant of the maser emission pumped by
  spherical winds in the AGB phase.
\item For the other objects previously cataloged as possible PNe
  (IRAS 17443$-$2949, IRAS 17560$-$3111, and IRAS 17579$-$3121), the
  positions of the 
  OH and/or H$_2$O maser emission and those of the most likely near-
  and mid-infrared counterparts of the IRAS sources are not consistent
  with the positions of the radio continuum emission reported in the
  literature. Therefore, these IRAS sources are not likely to trace
  PNe. 
\item We suggest an evolutionary scheme in which the precursors of
  H$_2$O-PNe would be ``water-fountain'' AGB or post-AGB stars. Water
  maser emission in these fountains tend to trace highly collimated
  jets (and equatorial structures in some cases). The jet clears the
  circumstellar envelope along the polar direction and, therefore,
  water masers in PNe are found preferentially tracing equatorial
  structures. 
\item Although IRAS 17443$-$2949, IRAS 17560$-$3111, and IRAS
  17579$-$3121 would be classified as OHPNe (objects with both OH
  maser and radio continuum emission, which have been suggested to be
  extremely young PNe) based on single-dish observations, our
  interferometric data
  indicate that they are not proper members of this class. Other
  prospective members of the OHPNe class will also need to be
  confirmed with interferometric observations, before drawing further
  conclusions about the properties of this class of objects.
\end{itemize}

\acknowledgments

JFG, JMT, GA, and OM acknowledge support from MEC (Spain) grant AYA
2005-08523-C03. LFM is supported by MEC grant AYA2005-01495.
OS is partially supported by MEC grant AYA2003-09499.
JFG, OS, LFM, JMT, and GA are also supported by Consejer\'{\i}a de
Innovaci\'on, Ciencia y Empresa of Junta de Andaluc\'{\i}a. YG is
supported by DGAPA-UNAM grant IN100407 and CONACyT grant 49947.
This publication makes use of data products from the Two
Micron All Sky Survey, which is a joint project of the University of
Massachusetts and the Infrared Processing and Analysis
Center/California Institute of Technology, funded by the National
Aeronautics and Space Administration and the National Science
Foundation.

\clearpage

\begin{deluxetable}{lrllrrrrlrr}
\rotate 
\tablecaption{Observational parameters: line data\label{tab.obslogline}
}
\tablewidth{0pt}
\tablehead{
\colhead{Molecule} & \colhead{$\nu_\circ$\tablenotemark{a}} & \colhead{Date\tablenotemark{b}} &
\colhead{Conf\tablenotemark{c}} & \colhead{FCal\tablenotemark{d}} & \colhead{$S_{\rm FCal}$\tablenotemark{e}} &
\colhead{PCal\tablenotemark{f}} & \colhead{$S_{\rm PCal}$\tablenotemark{e}} & \colhead{BW\tablenotemark{g}} &
\colhead{$N_c$\tablenotemark{h}} & \colhead{$\Delta V$\tablenotemark{i}} \\
 & \colhead{(MHz)} & & &  & \colhead{(Jy)} & & \colhead{(Jy)} &
 \colhead{(MHz)}  & & \colhead{(km s$^{-1}$)} 
}
\startdata
H$_2$O & 22235.08 & 2006-JAN-02 & D & J1331+305 & 2.54 & J1820-254 &
$0.744\pm 0.017$ & 3.125 & 64 & 0.66\\
H$_2$O & 22235.08 & 2006-OCT-06 & CnB & J1331+305 & 2.54 & J1820-254 &
$0.682\pm 0.022$ & 3.125 & 64 & 0.66\\
OH  & 1612.2310 & 2006-OCT-15 & CnB & J1331+305 & 13.85 & J1751-253 & $0.921\pm
0.017$ &  1.5625 & 256 & 1.14 \\
OH  & 1665.4018 & 2006-NOV-22 & C & J1331+305 & 13.63 & J1751-253 & $1.049\pm
0.007$ & 1.5625 & 128 & 2.20 \\
\enddata
\tablenotetext{a}{Rest frequency of the transition.}
\tablenotetext{b}{Date of observation. On 2006-JAN-02, only IRAS
  18061$-$2505 was observed. On 2006-OCT-06, IRAS 17443$-$2949 and
  IRAS 17580$-$3111 were observed. On the other two dates, all three
  sources were observed.}
\tablenotetext{c}{Configuration of the VLA.}
\tablenotetext{d}{Flux and bandpass calibrator}
\tablenotetext{e}{Flux density of calibrators.}
\tablenotetext{f}{Phase calibrator.}
\tablenotetext{g}{Bandwidth.}
\tablenotetext{h}{Number of observed spectral channels.}
\tablenotetext{i}{Velocity resolution.}
\end{deluxetable}

\begin{deluxetable}{rllrrrrl}
\rotate 
\tablecaption{Observational parameters: continuum data\label{tab.obslogcont}}
\tablewidth{0pt}
\tablehead{
\colhead{$\nu_{\rm sky}$\tablenotemark{a}} & \colhead{Date\tablenotemark{b}} &
\colhead{Conf\tablenotemark{c}} & \colhead{FCal\tablenotemark{d}} &
\colhead{$S_{\rm FCal}$\tablenotemark{e}} & 
\colhead{PCal\tablenotemark{f}} & \colhead{$S_{\rm
    PCal}$\tablenotemark{e}} & \colhead{BW\tablenotemark{g}}  \\
\colhead{(MHz)} & & &  & \colhead{(Jy)} & & \colhead{(Jy)} &
 \colhead{(MHz)} 
}
\startdata
22283 &  2006-JAN-02 & D & J1331+305 & 2.54 & J1820-254 & $0.729\pm
0.016$ & 25 \\
22283 &  2006-OCT-06 & CnB & J1331+305 & 2.54 & J1820-254 & $0.688\pm
0.016$ & 25\\
1685 & 2006-NOV-22\tablenotemark{h} & C & J1331+305 & 13.55 & J1751-253 & $1.18\phantom0\pm
0.04\phantom0$ & 25 \\
\enddata
\tablenotetext{a}{Central sky frequency of the observation.}
\tablenotetext{b}{Date of observation. On 2006-JAN-02, only IRAS
  18061$-$2505 was observed. On 2006-OCT-06, IRAS 17443$-$2949 and
  IRAS 17580$-$3111 were observed. On 2006-NOV-22, all three
  sources were observed.}
\tablenotetext{c}{Configuration of the VLA.}
\tablenotetext{d}{Flux and bandpass calibrator.}
\tablenotetext{e}{Flux density of calibrators.}
\tablenotetext{f}{Phase calibrator.}
\tablenotetext{g}{Bandwidth.}
\tablenotetext{h}{These data were not usable, due to radio frequency
  interference.} 
\end{deluxetable}

\begin{deluxetable}{llllrrr}
\tablecaption{Parameters of target sources and maps\label{tab.sourcedata}}
\tablewidth{0pt}
\tablehead{\colhead{IRAS name} & \colhead{Obs type\tablenotemark{a}} &
  \colhead{R.A.(J2000)\tablenotemark{b}} &
  \colhead{Dec(J2000)\tablenotemark{b}} &  \colhead{$V_{\rm LSR}$\tablenotemark{c}} &
  \colhead{$\theta_{\rm syn}$\tablenotemark{d}} & \colhead{p.a.\tablenotemark{e}}
  \\
& & & & (km s$^{-1}$) & (arcsec) & (degree)
}
\startdata
17443$-$2949 & OH-1612 & 17 47 35.30 & $-29$ 50 57.0 & $-5.0$ &
$13.0\times 8.0$ & $-78$ \\
             & OH-1665 & 17 47 35.30 & $-29$ 50 57.0 & $-5.0$ &
             $29.2\times 9.0$ & $-7$ \\
             & H$_2$O+cont & 17 47 35.30 & $-29$ 50 54.2 & $-5.0$ &
             $0.78\times 0.70$ & 72 \\ 
17580$-$3111 & OH-1612 & 18 01 19.60 & $-31$ 11 22.0 & 22.0 & $12.4\times
8.3$ & $80$ \\
             & OH-1665 & 18 01 19.60 & $-31$ 11 22.0 & 22.0 &
             $30.0\times 9.0$ & 2 \\
             & H$_2$O+cont  & 18 01 19.70 & $-31$ 11 23.0 & 22.0 & $1.1\times
             0.5$ & 44 \\ 
18061$-$2505 & OH-1612 & 18 09 12.40 &  $-25$ 04 34.0 & 60.0 &
$14.0\times 5.7$ & $65$\\
             & OH-1665 & 18 09 12.40 &  $-25$ 04 34.0 & 60.0 &
             $25.8\times 9.4$ & $13$  \\
             & H$_2$O+cont  & 18 09 12.40 &  $-25$ 04 34.0 & 60.0 & $5.3\times
             2.6$ & $-7$ 
\enddata
\tablenotetext{a}{Type of observation. OH-1612: OH maser line, rest
  frequency 1612 MHz; OH-1665: OH maser line, rest frequency 1665 MHz;
  H$_2$O+cont : Water maser line and continuum, simultaneous observations at
  22 GHz.}
\tablenotetext{b}{Coordinates of the phase center of the
  observations. Units of right ascension are
  hours, minutes, and seconds. Units of declination are degrees,
 arcminutes, and arcseconds.}
\tablenotetext{c}{Velocity of the bandwidth center with respect to the
  Local Standard of Rest (LSR), for line observations.}
\tablenotetext{d}{Full width at half maximum of the synthesized beam of
  the map.}
\tablenotetext{e}{Position angle (north to east) of the major axis of
  the synthesized beam.}
\end{deluxetable}

\begin{deluxetable}{llllrrr}
\rotate
\tablecaption{Detected maser components and upper limits of non-detections
\label{tab:result_line}}
\tablewidth{0pt}
\tablehead{
\colhead{IRAS} & \colhead{line} & \colhead{R.A.(J2000)} &
\colhead{Dec(J2000)} &  \colhead{$\Delta p$\tablenotemark{a}} & \colhead{$V_{\rm peak}$\tablenotemark{b}}
& \colhead{$S_{\rm  peak}$\tablenotemark{c}} \\
               &                   &                       &
                     &  \colhead{(milliarcsec)} & \colhead{(km s$^{-1}$)} 
& \colhead{(Jy)} 
}
\startdata
17442$-$2942 & OH-1612 & 17 47 28.867 & $-29$ 43 37.98 & $300\times 170$
& $-55.9$ & $0.229$ \\
             &      & 17 47 28.896 & $-29$ 43 38.41 & $400\times 250$
& $-24.1$ & $0.158$ \\
             & OH-1665 & \nodata & \nodata & \nodata & \nodata &
$\leq 0.010$\\
17443$-$2949 & OH-1612 & 17 47 35.22 & $-29$ 50 53.5 & $900\times 600$
& $-4.8$ & $0.056$ \\
             &      & 17 47 35.442 & $-29$ 50 53.32 & $40\times 30$
& $-16.1$ & $1.215$\\
             & OH-1665 & 17 47 35.16 & $-29$ 50 57.6 & $4000\times 1200$ 
& $2.6$ & $0.023$\\
             &      & 17 47 35.27 & $-29$ 50 52.0 & $3000\times 800$ &
$-10.6$ & $0.032$\\
             &      & 17 47 35.32 & $-29$ 50 54.1 & $1400\times 400$
& $-6.2$ & $0.063$\\
             &      & 17 47 35.326 & $-29$ 50 53.05 & $600\times 190$
& $-15.0$ & $0.141$\\
             & H$_2$O\tablenotemark{d} & 17 47 35.3253 & $-29$ 50 53.369 & $6 \times 5$
& $0.9$ & $0.690$\\ 
             &        & 17 47 35.32871 & $-29$ 50 53.3563 &
$0.8\times 0.7$ & $-3.7$ & $5.683$\\
             &        & 17 47 35.33063 & $-29$ 50 53.3984 & $0.6 \times
0.5$ & $-6.3$ & $6.912$ \\
            &        & 17 47 35.331 & $-29$ 50 53.34 & $50\times
40$ & $-1.7$ & $0.079$\\ 
17579$-$3121 & OH-1612\tablenotemark{d} & 18 01 13.3709 & $-31$ 21 56.596 & $10\times
7$ & $21.8$ & $4.504$ \\
             &      & 18 01 13.3771 & $-31$ 21 57.000 & $13\times
9$ & $1.4$ & $2.100$\\
             &      & 18 01 13.399 & $-31$ 21 57.38 & $190\times
130$ & $33.2$ & $0.134$\\
             & OH-1665 & \nodata & \nodata & \nodata & \nodata &
$\leq 0.010$\\
17580$-$3111 & OH-1612\tablenotemark{d} & 18 01 20.3987 & $-31$ 11 20.548 & $15\times
10$ & $28.6$ & $1.604$\\
             &      & 18 01 20.3992 & $-31$ 11 20.582 & $30\times
21$ & $12.7$ & $0.912$\\
             & OH-1665 & \nodata & \nodata & \nodata & \nodata &
$\leq 0.008$\\
             & H$_2$O & 18 01 20.3923 & $-31$ 11 21.305 & $21\times
11$ & $17.4$ & $0.418$\\
18061$-$2505 & OH-1612 & \nodata & \nodata & \nodata & \nodata &
$\leq 0.011$\\         
             & OH-1665 & \nodata & \nodata & \nodata & \nodata &
$\leq 0.007$\\
             & H$_2$O\tablenotemark{d} & 18 09 12.4022 & $-25$ 04 34.396 & $4\times 1.8$
& $57.4$ & $1.384$\\
             &        & 18 09 12.4032 & $-25$ 04 34.431 & $1.0\times 0.5$
& $60.7$ & $5.682$\\
             &        & 18 09 12.4033 & $-25$ 04 34.417 & $10\times 5$
& $63.9$ & $0.693$
\enddata
\tablenotetext{a}{Relative positional uncertainties within maps for
  each maser transition and source. Quoted uncertainties are 2$\sigma$
  errors, and have been
  estimated as $\Delta p = \theta_{\rm syn} \frac{\sigma_{\rm
      map}}{S_{\rm peak}}$, where  $\theta_{\rm syn}$ is the
  synthesized beam, and $\sigma_{\rm
      map}$ is the rms noise level of the map. The position angle of
    the error ellipse is the same as that of the synthesized beam. 
 Absolute astrometric errors can be estimated by quadratically adding
 these values to
 the intensity-independent uncertainties, roughly one tenth of the
 synthesized bean of each map.} 
\tablenotetext{b}{LSR velocity of the peak of each individual spectral
  component. Its uncertainty is determined by the velocity resolution of the
data.} 
\tablenotetext{c}{Peak flux density of individual spectral
  components, or upper limits for non-detections. Typical 2$\sigma$
  errors are 5-10 mJy. Upper limits are 3$\sigma$. Values have not been
  corrected by primary beam response.}
\tablenotetext{d}{Data were Hanning-smoothed in frequency. The resulting
  spectral resolution is double of their corresponding value in
  Table. \ref{tab.obslogline}.}
\end{deluxetable}

\begin{deluxetable}{rrllr}
\tablecaption{Continuum emission\label{tab:result_cont}}
\tablewidth{0pt}
\tablehead{
\colhead{IRAS} & \colhead{$\nu_{\rm sky}$}  &  \colhead{R.A.(J2000)} &
\colhead{Dec(J2000)} & \colhead{$S_\nu$} \\
               & \colhead{(MHz)} & & & \colhead{(mJy)}
}
\startdata
17442$-$2941 & 1665 & \nodata & \nodata & $\leq 0.9$\\
17443$-$2949 & 1665 & \nodata & \nodata & $\leq 0.9$\\
             & 22284  & \nodata & \nodata & $\leq 1.0$\\
17579$-$3121 & 1665 & \nodata & \nodata & $\leq 0.5$\\
17580$-$3111 & 1665 & \nodata & \nodata & $\leq 0.5$\\
             & 22282  & \nodata & \nodata & $\leq 1.4$\\
18061$-$2505 & 1665 & 18 09 12.563 & $-$25 04 32.51 &  $2.6\pm 0.9$\\
             & 22283 & 18 09 12.4036 & $-$25 04 34.383 & $38.1\pm 0.6$
\enddata
\end{deluxetable}

\begin{figure}
\includegraphics[scale=.6]{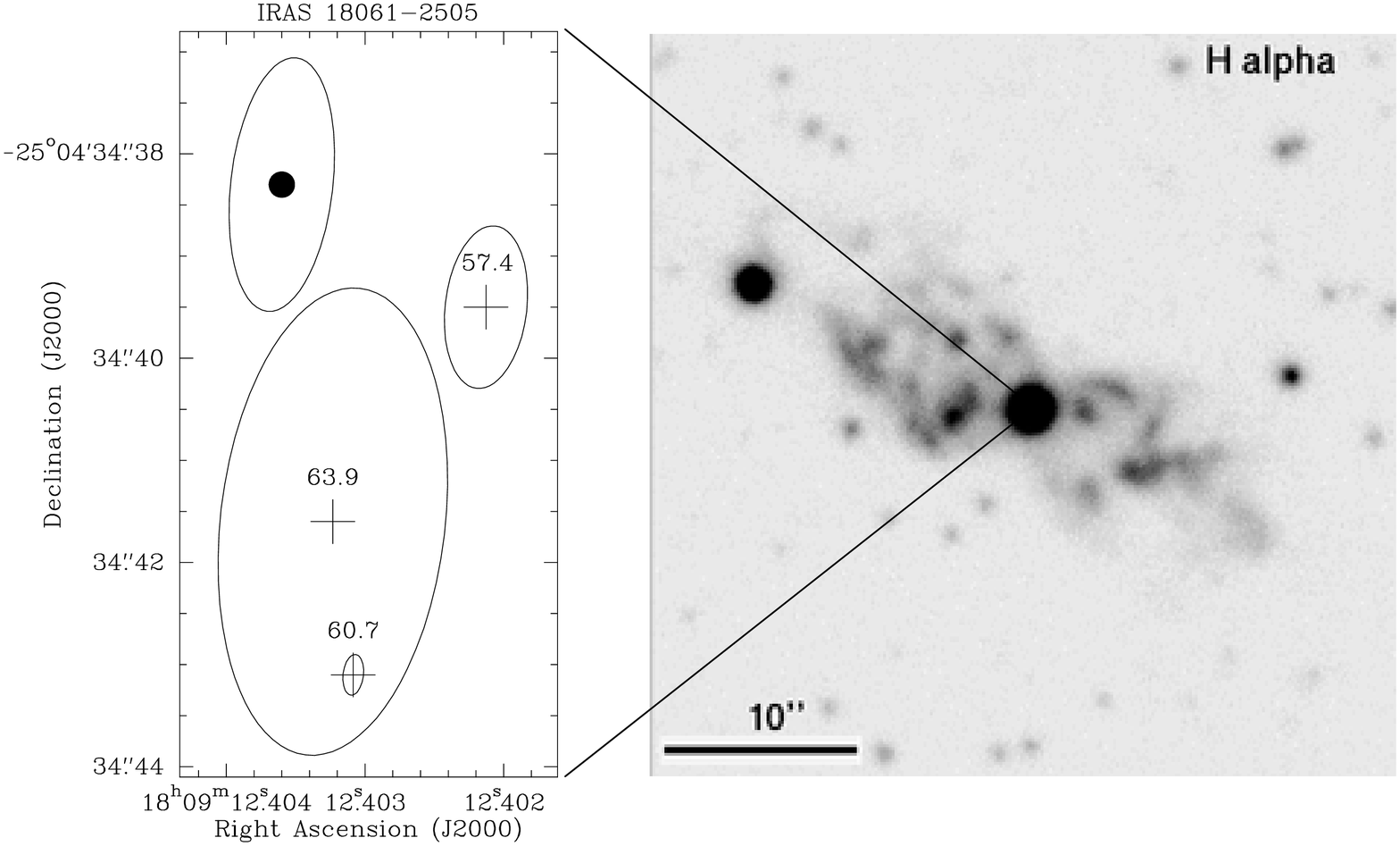}
\caption{Left: Positions of H$_2$O maser components (crosses) and 1.3 cm radio
  continuum emission (filled circle)
  in the IRAS 18061$-$2505 region. 
 Ellipses represent the {\em relative} positional errors within the
map. Absolute positional errors are $\simeq 0\farcs 5$.
The
  H$_2$O  components are labelled with their LSR velocities, in km s$^{-1}$. 
 Right: H$_\alpha$
image from \citet{Sua06}. Note the very different spatial scales in both
images. 
}
\label{fig:map18061all} 
\end{figure}

\begin{figure}
\includegraphics[scale=0.8,angle=-90]{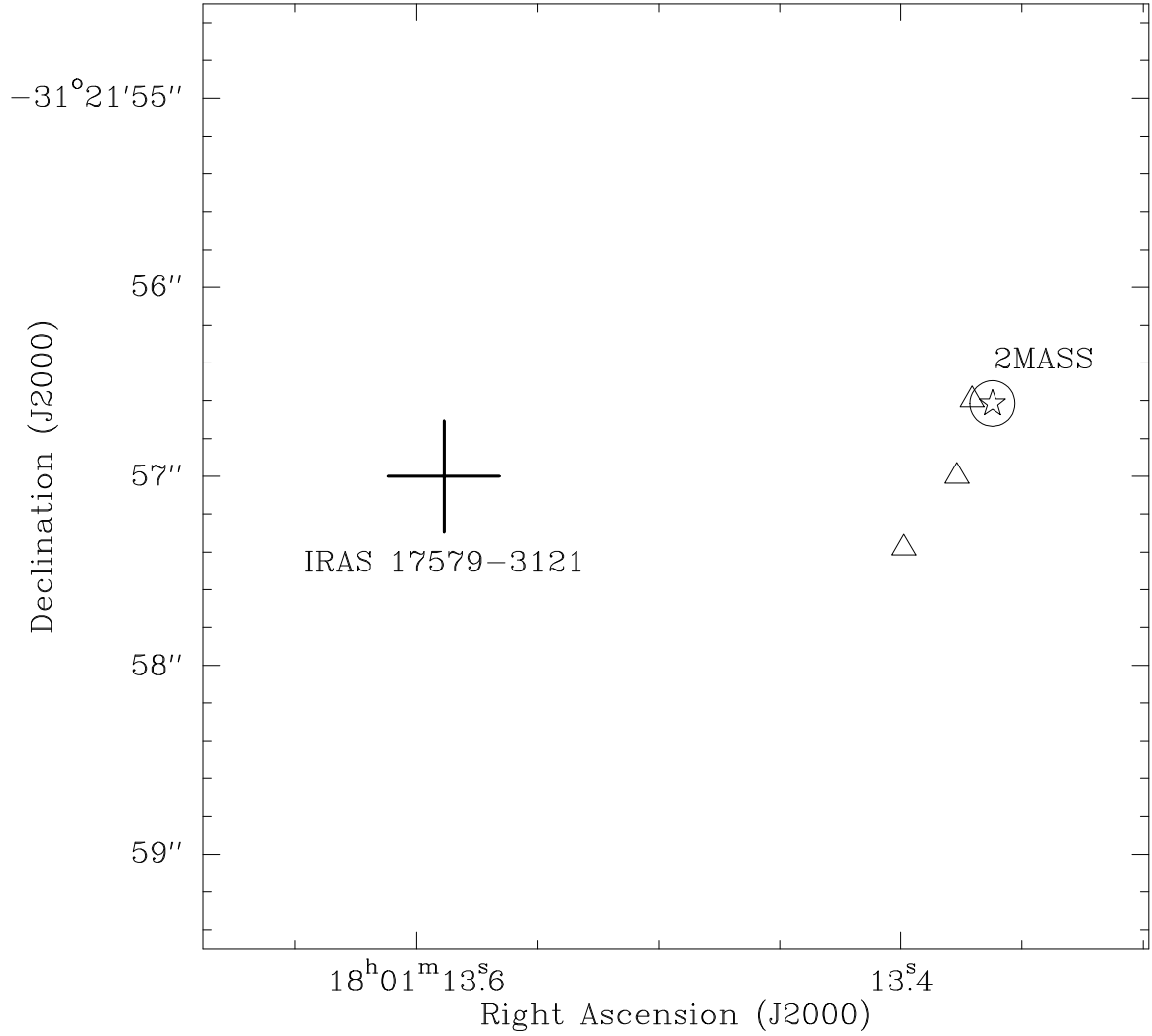}
\caption{Position of OH maser components and infrared sources in the IRAS
  17579$-$3121 field. Open triangles represent the OH masers at 1612 MHz. The
  five-pointed star marks the position of 2MASS J18011337-3121566. Absolute
  position errors of OH maser components are $\simeq 1"$. The large
  cross marks the position of the IRAS source (semiaxes
of the error ellipse are $39''\times 8''$, p.a. = $91^\circ$)} 
\label{fig:map17579all} 
\end{figure}

\end{document}